\documentclass[review,11pt]{elsarticle}

\usepackage{graphicx}
\usepackage{subfigure}
\graphicspath{{./figures/}}
\usepackage{amssymb} 
\usepackage{amsmath} 
\usepackage{physics}
\usepackage{xcolor}
\usepackage{multirow}
\usepackage{comment}
\usepackage[normalem]{ulem}
\usepackage{delimset}
\usepackage[margin=2.5cm]{geometry}
\usepackage[labelsep=period]{caption}

\usepackage{url}
\usepackage[version=4]{mhchem}
\usepackage[per-mode=symbol]{siunitx}

\usepackage{enumerate}
\usepackage{hyperref}
\hypersetup{colorlinks=true,linkcolor=blue,citecolor=blue, 
			bookmarksnumbered=true,bookmarksopen=true,
			pdfauthor=author,
			pdfpagelayout=SinglePage,pdfstartview=Fit}

\usepackage[capitalise,sort&compress]{cleveref}

\usepackage{booktabs}
\usepackage{multirow}
\usepackage{threeparttable}
\usepackage{setspace}   
\usepackage{array}
\usepackage{makecell}

\usepackage{framed}
\usepackage{multicol}
\usepackage{nomencl}
\makenomenclature
\printnomenclature[1.0cm]
\renewcommand*\nompreamble{\begin{multicols}{2}}
\renewcommand*\nompostamble{\end{multicols}}

\usepackage{microtype}
\usepackage[T1]{fontenc}
\usepackage[utf8]{inputenc}
\usepackage{libertine}
\usepackage[libertine]{newtxmath}
\usepackage{inconsolata}

\biboptions{numbers,sort&compress}
\begin{document}

\begin{frontmatter}
	\title{Diffusion-aware voltage source: An equivalent circuit network to resolve lithium concentration gradients in active particles}
	\author[ic,faraday]{Mingzhao Zhuo\corref{cor}}
	\ead{m.zhuo@imperial.ac.uk, mzhuo@connect.ust.hk}
	\author[ic]{Niall Kirkaldy}
	\ead{n.kirkaldy@imperial.ac.uk}
	\author[wae]{Tom Maull}
	\ead{Tom.Maull@wae.com}
	\author[wae]{Timothy Engstrom}
	\ead{Timothy.Engstrom@wae.com}
	\author[ic,faraday]{Gregory Offer}
	\ead{gregory.offer@imperial.ac.uk}
	\author[ic,faraday]{Monica Marinescu}
	\ead{monica.marinescu@imperial.ac.uk}
	\address[ic]{Department of Mechanical Engineering, Imperial College London, London SW7 2AZ, United Kingdom}
	\address[wae]{WAE Technologies, Wantage OX12 0DQ, United Kingdom}
	\address[faraday]{The Faraday Institution, Didcot OX11 0RA, United Kingdom}
	\cortext[cor]{Corresponding author}
	\begin{abstract}
		Traditional equivalent circuit models (ECMs) have difficulties in estimating battery internal states due to the lack of relevant physics, such as the lithium diffusion in active particles.
		Here we configure a circuit network to describe the lithium diffusion and define it as a new high-level circuit element called diffusion-aware voltage source. 
		The circuit representation is proven equivalent to the discretized diffusion equation. 
		The new voltage source gives the electrode potential as a function of the surface concentration and thus automatically incorporates the diffusion overpotential. 
		We show that an ECM with the proposed diffusion-aware voltage sources (called ``shell ECM'') can reproduce the single particle model simulation results, making it a trustworthy easy-to-implement substitute. 
		Furthermore, the simplest shell ECM consisting of a single diffusion-aware voltage source and a resistor is validated against experimental constant-current discharges at various rates. 
		The diffusion-aware voltage source can be used to measure diffusivity by fitting the diffusion resistance against experimental data. 
		The viability of the shell ECM for onboard usage is confirmed by implementation into a battery management system of WAE Technologies. 
		By tracking the internal concentration states, the shell ECM demonstrates robustness to dynamic applied-current profiles. 
	\end{abstract}


\end{frontmatter}


\begin{table*}
	\begin{framed}
	\small

		\renewcommand{\nomname}{Acronyms}
		\begin{thenomenclature}

			\nomgroup{A}

			\item [{ECM}]\begingroup equivalent circuit models \nomeqref {0}
			\nompageref{1}

			\item [{PE}]\begingroup positive electrode \nomeqref {0}
			\nompageref{1}

			\item [{OCP}]\begingroup electrode open circuit potential \nomeqref {0}
			\nompageref{1}

			\item [{DFN}]\begingroup Doyle-Fuller-Newman (or called P2D) \nomeqref {0}
			\nompageref{1}

			\item [{GITT}]\begingroup Galvanostatic Intermittent Titration Technique \nomeqref {0}
			\nompageref{1}

			\item [{SoC}]\begingroup state of charge \nomeqref {0}
			\nompageref{1}


			\item [{RC}]\begingroup resistor capacitor \nomeqref {0}
			\nompageref{1}

			\item [{NE}]\begingroup negative electrode \nomeqref {0}
			\nompageref{1}

			\item [{OCV}]\begingroup cell open circuit voltage \nomeqref {0}
			\nompageref{1}

			\item [{SPM}]\begingroup single particle model \nomeqref {0}
			\nompageref{1}

			\item [{RMSE}]\begingroup root mean square error \nomeqref {0}
			\nompageref{1}

			\item [{BMS}]\begingroup battery management system \nomeqref {0}
			\nompageref{1}

		\end{thenomenclature}

		\renewcommand{\nomname}{Nomenclature}
		\begin{thenomenclature}

			\nomgroup{A}

			\item [{$c$}]\begingroup concentration variable in active particles \nomeqref {0}
			\nompageref{1}

			\item [{$c_n$}]\begingroup average concentration in layer $n$ \nomeqref {0}
			\nompageref{1}

			\item [{$c_{\text{p}}$}]\begingroup lithium concentration in PE particles \nomeqref {0}
			\nompageref{1}

			\item [{$c_{\text{n}}$}]\begingroup lithium concentration in NE particles \nomeqref {0}
			\nompageref{1}

			\item [{$c_{\text{s,surf}}$}]\begingroup particle-surface concentration \nomeqref {0}
			\nompageref{1}

			\item [{$c_{\text{s,max}}$}]\begingroup maximum concentration in active particles \nomeqref {0}
			\nompageref{1}

			\item [{$D_\text{s}$}]\begingroup lithium diffusivity in active particles \nomeqref {0}
			\nompageref{1}

			\item [{$N$}]\begingroup maximum number of layers \nomeqref {0}
			\nompageref{1}

			\item [{$n$}]\begingroup layer number ranging from 1 to $N$ \nomeqref {0}
			\nompageref{1}

			\item [{$a$}]\begingroup active-particle radius \nomeqref {0}
			\nompageref{1}

			\item [{$b$}]\begingroup thickness of each layer \nomeqref {0}
			\nompageref{1}

			\item [{$\Omega_n$}]\begingroup volume of layer $n$ \nomeqref {0}
			\nompageref{1}

			\item [{$S_n$}]\begingroup outermost surface area of layer $n$ \nomeqref {0}
			\nompageref{1}

			\item [{$J_n$}]\begingroup lithium mass flux from layer $n$ to $n+1$ \nomeqref {0}
			\nompageref{1}

			\item [{$\bar{I}_n$}]\begingroup current flux from layer $n$ to $n+1$ \nomeqref {0}
			\nompageref{1}

			\item [{$R_{\text{d},n}$}]\begingroup diffusion resistance between layer $n$ and $n+1$ \nomeqref {0}
			\nompageref{1}

			\item [{$V_n$}]\begingroup voltage of voltage source $n$ \nomeqref {0}
			\nompageref{1}

			\item [{$V_\text{t}$}]\begingroup cell terminal voltage \nomeqref {0}
			\nompageref{1}

			\item [{$I_n$}]\begingroup current flowing into voltage source $n$ \nomeqref {0}
			\nompageref{1}

			\item [{$k$}]\begingroup constant coefficient for unit conversion \nomeqref {0}
			\nompageref{1}
			
			\item [{$I_\text{app}$}]\begingroup applied current \nomeqref {0}
			\nompageref{1}

			\item [{$I_\text{in}$}]\begingroup applied current taken by one particle \nomeqref {0}
			\nompageref{1}

			\item [{$\tau$}]\begingroup diffusion timescale \nomeqref {0}
			\nompageref{1}

			\item [{$R_\text{ct}$}]\begingroup charge-transfer resistance \nomeqref {0}
			\nompageref{1}

			\item [{$C_\text{dl}$}]\begingroup double-layer capacitance \nomeqref {0}
			\nompageref{1}

			\item [{$R_\text{0}$}]\begingroup resistance for instantaneous voltage change \nomeqref {0}
			\nompageref{1}

			\item [{$R_\text{s}$}]\begingroup resistance for electronic conduction in the solid \nomeqref {0}
			\nompageref{1}

			\item [{$R_\text{e}$}]\begingroup electrolyte resistance \nomeqref {0}
			\nompageref{1}

			\item [{$\epsilon_\text{a}$}]\begingroup volume fraction of active materials \nomeqref {0}
			\nompageref{1}

			\item [{$A$}]\begingroup total surface area of active materials \nomeqref {0}
			\nompageref{1} 

			\item [{$V$}]\begingroup electrode volume \nomeqref {0}
			\nompageref{1}

			\item [{$N_\text{a}$}]\begingroup number of active particles \nomeqref {0}
			\nompageref{1}

			\item [{$Q$}]\begingroup cell nominal capacity \nomeqref {0}
			\nompageref{1}

			\item [{$j$}]\begingroup interfacial current density \nomeqref {0}
			\nompageref{1}

			\item [{$j_\text{0}$}]\begingroup exchange current density \nomeqref {0}
			\nompageref{1}

			\item [{$m$}]\begingroup charge-transfer reaction rate \nomeqref {0}
			\nompageref{1}

			\item [{$\eta$}]\begingroup reaction overpotential \nomeqref {0}
			\nompageref{1}

			\item [{$S_\text{a}$}]\begingroup particle surface area to volume ratio \nomeqref {0}
			\nompageref{1}

			\item [{$ F $}]\begingroup Faraday constant \nomeqref {0}
			\nompageref{1}

			\item [{$ R $}]\begingroup gas constant \nomeqref {0}
			\nompageref{1}

			\item [{$ T $}]\begingroup absolute temperature \nomeqref {0}
			\nompageref{1}

		\end{thenomenclature}

	\end{framed}
\end{table*}

%
\section{Introduction}

During (dis)charge of lithium-ion batteries, one of the main limiting physical processes is the slow diffusion of lithium in active material particles~\cite{plett2015battery}, impacting battery fast-charge performance~\cite{Weiss2021}. 
This mechanism has been well considered in physics-based models such as the widely-used Doyle-Fuller-Newman (DFN) model~\cite{Doyle1993}; however, it is not sufficiently described in traditional equivalent circuit models (ECMs). 
Traditional ECMs account for the diffusion-induced overpotential by a series of resistor-capacitor (RC) pairs but cannot capture the internal concentration states, especially the particle surface concentration for state of available power (SoAP) estimation.
Here, we propose a circuit network consisting of diffusion resistors and voltage sources to describe the solid diffusion process and resolve lithium concentration distribution to assist internal state estimation and battery control.

The fundamental physical processes in lithium-ion batteries include lithium diffusion in active materials, lithium-ion diffusion and migration in the electrolyte, electronic conduction in the solid conductive materials, and intercalation chemical reactions at the solid-electrolyte interface. 
These processes are well understood and described in the DFN model by partial differential algebraic equations (PDAEs), leading to accurate prediction of battery behavior and estimation of internal states.
Besides these basic physical processes, other physics such as thermal behavior and battery ageing from side reactions have also been modeled by mathematical equations added to the DFN model. 
However, discretizing and solving the coupled PDAEs for battery composite materials are costly in terms of implementation and computation, and thus physics-based models are often infeasible for real-time application in state-of-art battery control systems.

In contrast, ECMs leverage well-established knowledge of electrical circuits (e.g., resistors, capacitors, and voltage sources) to mimic the cell behavior, in place of modelling the underlying physics.
The standard electrical elements allow the ECMs to be solved in real time so that ECMs are widely used in battery management systems (BMS) for control purpose.
However, the loss of connection to physical processes also hinders ECMs in internal state estimation and weakens their prediction power.
For example, the electrode open circuit potential (OCP) is a function of the surface concentration, and both the electrode OCP and surface concentration change during the cell rest following a charge/discharge due to the solid diffusion. 
Traditional ECMs use a Warburg resistance \cite{plett2015battery} to represent the solid diffusion effect and then approximate it with multiple (usually two) RC pairs. 
Although the RC pair has a characteristic time constant and can model the voltage change, it is still not aware of the concentration distribution in the particle and on the surface. 

The surface concentration is different from the average concentration as the slow diffusion leads to concentration gradients, especially under high dis(charge) rates. 
It cannot be directly measured or estimated from the measured current and voltage data. 
However, it is an important internal state in model-based battery control, for example, in the design of fast-charge protocols \cite{Yin2019} and in accurate SoAP estimation~\cite{chaturvedi2010alg,Zheng2018} for fast-discharge applications.

Continuous efforts \cite{Rael2013,Scipioni2017,von_Srbik_2016,Merla2018a,Sato2019,Li2019a,Geng2021} have been made in bringing physics into circuit-based models for better state estimation and more accurate prediction. 
The common idea is to use electrical analogies to represent the physical processes, including ageing mechanisms.
Most of these works focused on converting the DFN model into an equivalent circuit-based model, addressing the current distribution and charge transport processes.
The solid diffusion effect was addressed quite differently among the existing studies.
The first approach~\cite{Sato2019,Geng2021} is to solve the continuous diffusion equation using numerical techniques such as finite difference method and then use the solved surface concentration to calculate the electrode potential. 
Alternatively, Li et al.~\cite{Li2019a} used a two-parameter polynomial approximation to the numerical solution to obtain the surface potential, which is a common approximation technique for single particle models \cite{Subramanian2005}.
Furthermore, Ouyang et al.~\cite{Ouyang2014} and Zheng et al.~\cite{Zheng2019} disregarded the concentration distribution, but estimated the surface concentration from available approximate solutions to the diffusion equation.
A circuit-style approach is to use circuit networks comprising of resistors and capacitors \cite{Rael2013,von_Srbik_2016} or of resistors and voltage sources \cite{Merla2018a} to mimic the solid diffusion process. 
However, transforming the diffusion process from physical description to circuit arrangement has still not been addressed in a satisfactory way. 
The numerical solutions \cite{Sato2019,Geng2021} or approximate solutions~\cite{Li2019a,Ouyang2014,Zheng2019} appear to have the same level of implementation complexity as physics-based models. 
The circuit networks for solid diffusion in Merla et al.~\cite{Merla2018a} used the electrode OCP, rather than concentration gradients, as the driving force for diffusion. 
The resistor-capacitor networks \cite{Rael2013,von_Srbik_2016} proposed promising analogies but needed further improvement. 

A well-designed circuit representation implementing the correct solid diffusion physics is needed for ECMs to be easily parameterized and implemented in a real-world BMS. 
To this end, we propose a circuit network (\cref{sec:circuitnetwork}), consisting of controlled voltage sources and diffusion resistors, and define it as a new circuit element called diffusion-aware voltage source (\cref{fig:fullECM}). 
An ECM involving the diffusion-aware voltage sources, called ``shell ECM'' (\cref{sec:shellecm}), is configured to predict cell responses.
The proposed diffusion-aware voltage source and shell ECMs are verified (\cref{sec:compspm}) against the physics-based single particle model run in an open-source battery-modelling toolbox PyBaMM \cite{Sulzer2021}. 
The shell ECM is further validated against experimental data in \cref{sec:expval}. 
Finally, the shell ECM is implemented into the BMS of WAE Technologies and compared with traditional ECMs in terms of computational efficiency and accuracy (\cref{sec:imple}). 

%

\section{Equivalent circuit network for solid diffusion}

In this section, we present an equivalent circuit network for solid diffusion in active particles and then define the proposed equivalent circuit network as a diffusion-aware voltage source that can be used to configure physics-informed ECMs.
We start with a brief introduction to the continuous description of the solid diffusion and its spatial discretization by finite volume method in \cref{sec:diffusionPDE}.
Based on the spatially discretized diffusion equation, we devise the diffusion-aware voltage source (\cref{sec:circuitnetwork}) and propose battery-cell ECMs with the new voltage sources (\cref{sec:shellecm}), hereafter called shell ECMs.

\subsection{Preliminaries: diffusion equation and its discretization}
\label{sec:diffusionPDE}

\begin{figure}
	\centering
	\includegraphics{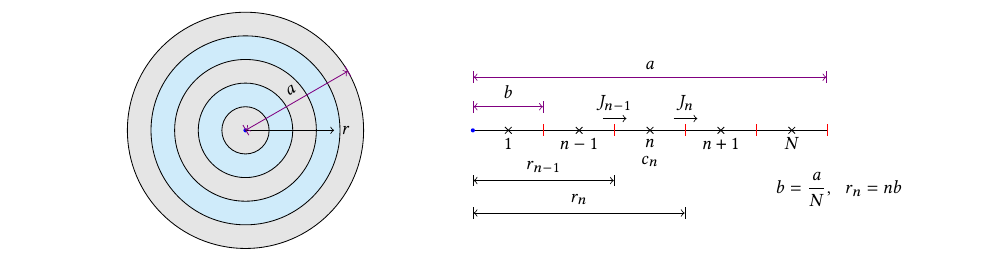}
	\caption{
		Schematics of discretizing an active particle of radius $a$.
		The particle is discretized into $N$ layers with layer number $n$ ranging from $1$ to $N$.
		The symbol $J_n$ denotes mass flux from layer $n$ to $n+1$, and $c_n$ is the lithium concentration in layer $n$.
	}
	\label{fig:discretization}
\end{figure}

The transport of intercalated lithium in active particles is often described, under the assumption of Fickian diffusion~\cite{BrosaPlanella2022}, by a partial differential equation that is expressed as
\begin{align} \label{eq:soliddiff}
	\pdv{c}{t} + \div{\qty(- D_{\text{s}} \grad{c})} & = 0,
\end{align}
where $c$ is the field variable of lithium concentration and $D_{\text{s}}$ is the diffusivity coefficient. 
Assuming spherical active particles, we reduce by symmetry the diffusion equation~(\ref{eq:soliddiff}) to
\begin{align} \label{eq:soliddiff_sph}
	\pdv{c}{t} + \frac{1}{r^2} \frac{\partial}{\partial r} \qty(-r^2 D_{\text{s}} \frac{\partial c}{\partial r}) = 0,
\end{align}
where $r$ denotes the radial direction (\cref{fig:discretization}) in a spherical coordinate system. 
The boundary conditions at the particle center and surface are, respectively, 
\begin{align} \label{eq:bcs}
	\left. -D_\text{s} \frac{\partial c}{\partial r} \right|_{r=0} = 0 \quad \text{and} \quad
	\left. -D_\text{s} \frac{\partial c}{\partial r} \right|_{r=a} = j/F,
\end{align}
where $j$ is the interfacial current density flowing out the particle and $a$ is the particle radius.

Following the finite volume method, we divide the sphere into $N$ layers of equal thickness, as shown in \cref{fig:discretization}, and integrate \cref{eq:soliddiff_sph} over an arbitrary layer $n$ (control volume):
\begin{align} \label{eq:integralcons}
	\int_{r_{n-1}}^{r_n} 4\pi r^2 \qty[\pdv{c}{t} + \frac{1}{r^2} \frac{\partial}{\partial r} \qty(-r^2 D_{\text{s}} \frac{\partial c}{\partial r})] \dd{r} = 0.
\end{align}
The concentration variable $c$ in layer $n$ is considered constant and equal to the average concentration denoted as $c_n$. 
We thus have
\begin{align} \label{eq:integradiff}
	\pdv{c_n}{t} \Omega_n + S_{n} J_{n} - S_{n-1}J_{n-1} = 0,
\end{align}
where $\Omega_n$ represents the volume of layer $n$, $S_{n}$ is the outermost surface area of layer $n$, and $J_n$ denotes the flux at the outermost surface: 
\begin{align} \label{eq:disclayern}
	\Omega_n = \frac{4}{3} \pi \qty(r_n^3 - r_{n-1}^3), \quad
	S_{n}=4\pi r_n^2, \quad
	J_n = -D_{\text{s}}\pdv{c}{r} = -D_\text{s} \frac{c_{n+1} - c_{n}}{b},
\end{align}
where $b$ is the thickness of each layer. 
Inserting \cref{eq:disclayern} into \cref{eq:integradiff}, we obtain the mass conservation for an arbitrary layer $n$:
\begin{align} \label{eq:discretized}
	\frac{\dd{c_n}}{\dd{t}} \Omega_n =
	\frac{c_{n+1} - c_{n}}{b/\qty(S_{n}D_\text{s})} - \frac{c_n - c_{n-1}}{b/\qty(S_{n-1}D_\text{s})}.
\end{align}

\subsection{Equivalent circuit representation}
\label{sec:circuitnetwork}

\begin{figure}
	\centering
	\includegraphics{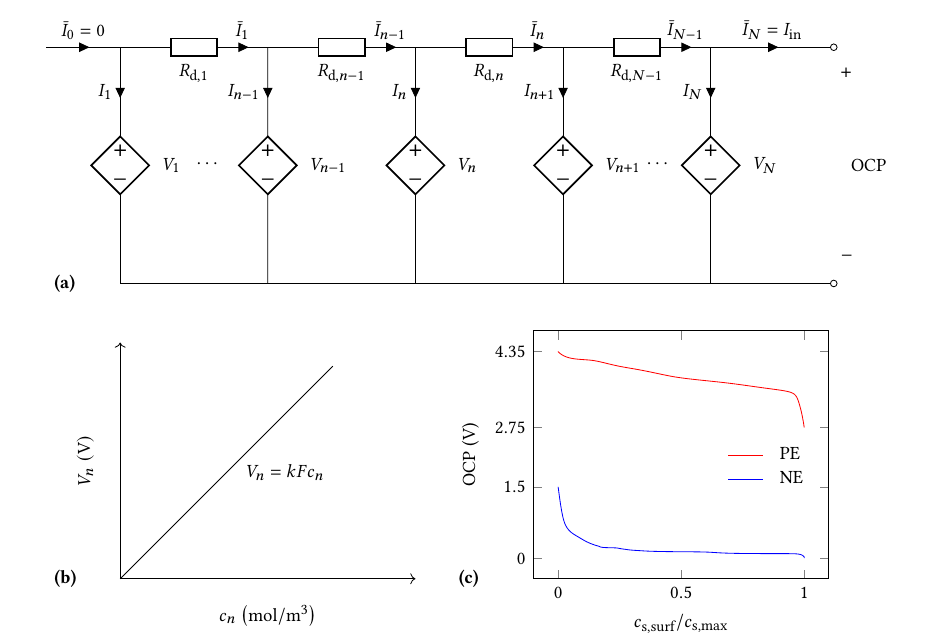}
	\caption{
		Schematics of shell circuit network (a) for solid diffusion description.
		The voltage $V_n$ ($n=1,2,\ldots,N$) of an arbitrary layer $n$ is a linear function of its internal state---lithium concentration $c_n$, as shown in (b).
		The open circuit potential (OCP) of the particle depends on the surface concentration normalized by the maximum concentration $c_\text{s,surf}/c_\text{s,max}$. 
		The real OCP relations for the positive electrode (PE) and negative electrode (NE) of a LG\,M50 cell \cite{Chen2020} are shown in (c) as an example.
	}
	\label{fig:multivolts}
\end{figure}

In traditional ECMs, solid diffusion in active particles is modeled by a Warburg element and a finite number of RC pairs~\cite{plett2015battery}, which can mimic the voltage response of a cell under certain conditions but cannot resolve the concentration gradient inside a particle. 
In particular, the surface concentration and real electrode potential cannot be obtained for improved state estimation. 
Here we use electrical analogies to describe the solid diffusion and configure voltage sources and resistors in a shell structure as shown in \cref{fig:multivolts}, in place of the differential equation description \cref{eq:soliddiff_sph}.

In the shell circuit network (\cref{fig:multivolts}), we have two electrical elements: resistors and controlled voltage sources.
The voltage source $V_n$ corresponds to the layer $n$ in \cref{fig:discretization} on a one-to-one basis. 
According to Kirchhoff's voltage law, for an arbitrary circuit loop in \cref{fig:multivolts}a we have
\begin{align} \label{eq:kvl}
	V_{n} = V_{n+1} + \bar{I}_{n} R_{\text{d},n}.
\end{align}
Kirchhoff's current law states that
\begin{align} \label{eq:kcl}
	I_n & = \bar{I}_{n-1} -  \bar{I}_{n}.
\end{align}
Calculating $\bar{I}_{n-1}$ and $\bar{I}_{n}$ from \cref{eq:kvl} and inserting them into \cref{eq:kcl}, we have
\begin{align} \label{eq:circuitgov}
	I_n & = \frac{V_{n-1} - V_{n}}{R_{\text{d},n-1}} - \frac{V_{n} - V_{n+1}}{R_{\text{d},n}}.
\end{align}
For each layer $n$, we further attach an internal state variable $c_n$ quantifying the local lithium concentration to the controlled voltage source $V_{n}$ and define the controlled voltage as 
\begin{align} \label{eq:controlledvolt}
	V_{n} = kFc_{n},
\end{align}
where $F$ is Faraday's constant and $k$ (\si{\volt\cubic\meter\per\coulomb}) is a constant coefficient to make units consistent in the left- and right-hand sides of \cref{eq:controlledvolt}. 
The value of $k$ has no impact and thus is set to be 1 hereafter. 
Substituting \cref{eq:controlledvolt} into \cref{eq:circuitgov}, we have
\begin{align} \label{eq:shelleq}
	\frac{I_n}{F} & = \frac{c_{n+1} - c_{n}}{R_{\text{d},n} / k} - \frac{c_{n} - c_{n-1}}{R_{\text{d},n-1} / k}.
\end{align}
The current flow $I_n$ in \cref{fig:multivolts} means charge flux into voltage source $V_n$, and this is analogous to the lithium mass flux into layer $n$. 
To draw the analogy, we use Faraday's constant $F$ to make the units consistent: 
\begin{align} \label{eq:connection}
	\frac{I_n}{F} = \frac{\dd{c_n}}{\dd{t}} \Omega_n,
\end{align}
indicating that the left-hand sides of \cref{eq:discretized} and \cref{eq:shelleq} are equal.
We can further define the diffusion resistances in the shell circuit network as
\begin{align} \label{eq:resistance}
	R_{\text{d},n} = \frac{kb}{S_{n}D_\text{s}} & = \frac{kb}{4\pi (nb)^2 D_\text{s}}
	= \frac{kN}{4\pi n^2 a D_\text{s}}. 
\end{align}
Now \cref{eq:discretized,eq:shelleq} are completely equivalent.
Therefore, with the definition of controlled voltage source in \cref{eq:controlledvolt} and diffusion resistance in \cref{eq:resistance}, the proposed shell circuit network is equivalent to the spatially discretized diffusion equation.
We remark that the driving force of current flow is the voltage difference and this reflects that the concentration gradient is the driving force for diffusion. 
The governing equations for the shell circuit network are summarized in \cref{tab:goveeqs}, and the numerical techniques to solve them can be found in \ref{app:num}. 
\begin{table}
	\centering
	\caption{Governing equations for the diffusion-aware voltage source.} \label{tab:goveeqs}
	\begin{threeparttable}
		\begin{tabular}{@{}l@{}}
		\toprule
			\begin{minipage}{1.0\textwidth}
				\vspace{0.3em}
				1. Kirchhof’s voltage law:
				\begin{align*} \tag{\ref{eq:kvl}}
					V_{n} = V_{n+1} + \bar{I}_{n} R_{\text{d},n}, \quad n=1,2,\ldots,N-1
				\end{align*}
				2. Kirchhof’s current law:
				\begin{align*} \tag{\ref{eq:kcl}}
					I_n = \bar{I}_{n-1} - \bar{I}_{n}, \quad n=1,2,\ldots,N
				\end{align*}
				3. Internal state evolution:
				\begin{align*} \tag{\ref{eq:connection}}
					\frac{\dd{c_n}}{\dd{t}} = \frac{1}{F\Omega_n} I_n \qty(t), \quad n=1,2,\ldots,N
				\end{align*}
				4. Constitutive relation: 
				\begin{align*} \tag{\ref{eq:controlledvolt}}
					V_n = k F c_n, \quad n=1,2,\ldots,N
				\end{align*}
				\vspace{0.1em}
			\end{minipage} \\
			\bottomrule
		\end{tabular}
	\end{threeparttable}
\end{table}

The integral form of the internal state $c_{n}$ of layer $n$ can be derived from \cref{eq:connection}:
\begin{align}
	c_{n} = c_{n,0} + \frac{1}{F\Omega_n} \int_{t_{0}}^{t} I_{n} \dd{t},
\end{align}
where $c_{n,0}$ represents the initial state, corresponding to the initial condition of diffusion equation~(\ref{eq:soliddiff_sph}) that is omitted for brevity. 
Finally, the two current flows at the edges of the shell circuit network are specified as
\begin{align}
	\bar{I}_0 = 0 \quad \text{and} \quad
	\bar{I}_{N} = I_\text{in},
\end{align}
where $I_\text{in}$ is the portion of applied current $I_\text{app}$ for one active particle (detailed below).
These two expressions correspond to the boundary conditions (\ref{eq:bcs}) of the diffusion equation.

The shell circuit network as a whole can be considered as a high-level circuit element whose input is the applied current.
The output is the electrode potential
\begin{align}
	\text{OCP} = \text{OCP}\qty(c_\text{s,surf}),
\end{align}
where $\text{OCP}$ is the measured potential function for an active material (see \cref{fig:multivolts}c for examples) and the surface concentration is linearly extrapolated from those of the outermost two layers
\begin{align} \label{eq:surfconcen}
	c_\text{s,surf} = 1.5c_{N} - 0.5c_{N-1}.
\end{align}

In \cref{fig:multivolts}, there are $N-1$ diffusion resistors, but only one resistance is unknown and needs to be determined. 
According to \cref{eq:resistance}, the innermost resistance is the largest and thus is chosen as the unknown and the baseline for calculating other diffusion resistances. 
It can be expressed as
\begin{align} \label{eq:outerres}
	R_{\text{d},1} = \frac{kN}{4\pi a D_\text{s}}. 
\end{align}
We remark that the diffusion resistance is inversely proportional to the solid diffusivity, and this relation can be used, as an alternative to the current approach~\cite{ORegan2022}, to measure the diffusivity of active materials in the laboratory (\ref{app:fittimescale}), being the diffusion resistance fitted from experimental data.
The shell circuit network is designed for the physical process of diffusion, and thus there must be a timescale associated with it. 
We define the timescale of the shell circuit network as that associated with the continuous diffusion equation: 
\begin{align} \label{eq:timescale}
	\tau = \frac{a^2}{D_\text{s}} = \frac{4\pi a^3 R_{\text{d},1}}{kN},
\end{align}
where the diffusivity $D_\text{s}$ is expressed in terms of diffusion resistance $R_{\text{d},1}$ from \cref{eq:outerres}.

\subsection{Shell equivalent circuit model}
\label{sec:shellecm}

\begin{figure}
	\centering
	\includegraphics{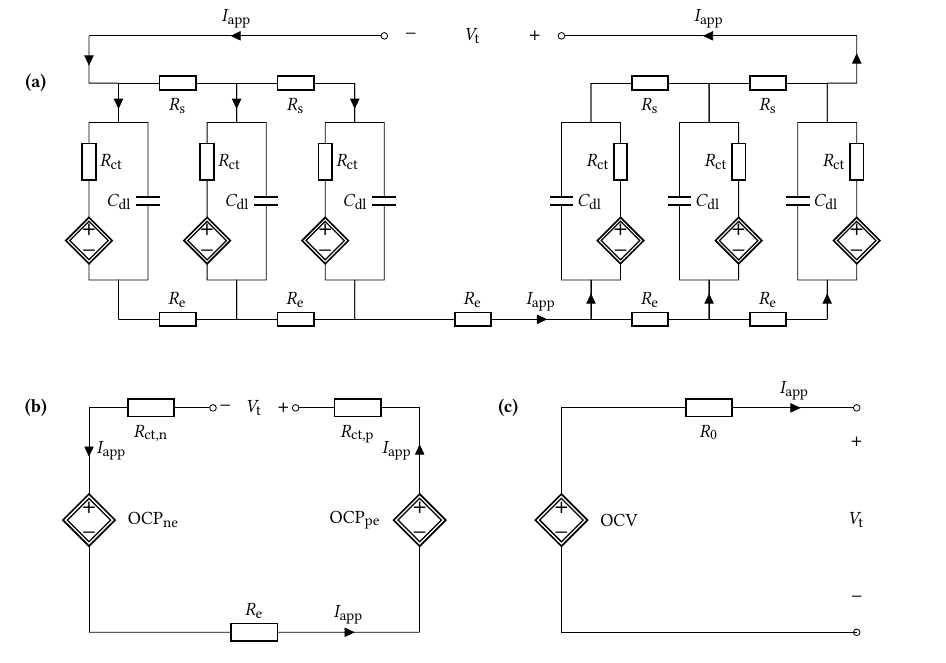}
	\caption{
		Shell ECMs equivalent to physics-based models---the DFN model (a) and Single Particle Model (b).
		Further simplified shell ECM (c) for easy experimental parameterization.
		The diffusion-aware voltage source (double-wall diamond symbol) represents the proposed shell circuit network in \cref{fig:multivolts}.
	}
	\label{fig:fullECM}
\end{figure}

In the previous section, we devise a circuit network for the solid diffusion.
This network can be considered as a new high-level circuit element that takes current as the input, stores internal states of concentration (or local SoC), and gives surface OCP as the output.
Thus, we define it as a diffusion-aware voltage source symbolized by a double-wall diamond as shown in \cref{fig:fullECM}.

Based on the diffusion-aware voltage source, we propose an ECM in \cref{fig:fullECM}a to resolve the fundamental physical processes described in the DFN model (detailed in Introduction) and call it the comprehensive shell ECM.
The comprehensive shell ECM consists of two similar sub-circuits in the left-hand and right-hand sides, corresponding to the negative electrode (NE) and positive electrode (PE), respectively.
The upper resistors $R_\text{s}$ are used to model the electrical potential variation in the electrode thickness direction due to the electronic current flow in solid conductive materials.
The lower resistors $R_\text{e}$ are designed for the overpotential caused by lithium-ion transport in the electrolyte.
Connecting electronic and ionic current flows is the intercalation and deintercalation chemical reactions on the active particle surface, represented by an elementary circuit unit involving a double layer capacitor $C_\text{dl}$, a charge transfer resistance $R_\text{ct}$, and the diffusion-aware voltage source.
This elementary circuit unit resembles the Randles circuit \cite{plett2015battery} but replaces the Warburg impedance with the newly defined diffusion-aware voltage source. 

In \cref{fig:fullECM}a, we only show for demonstration three elementary circuit units in parallel at each side; the more elementary units, the higher the resolution of current distribution and lithium concentration gradient across the electrode thickness.
The comprehensive shell ECM is thus especially suitable for high-current (dis)charges. 
For better interpretation of the diffusion-aware voltage source, we simplify the shell ECM to the extent shown in \cref{fig:fullECM}b.
We first disregard $R_\text{s}$ and $R_\text{e}$ as the electronic and ionic conduction are fast and the corresponding overpotentials are relatively small~\cite{Geng2021}. 
Then, we omit the double-layer capacitance as it has very little impact except at very high frequencies \cite{plett2015battery}. 
Consequently, the three elementary circuit units within each electrode would behave identically and thus can be integrated into one.

We remark that the simplification from (a) to (b) is based on the assumption that all particles in different locations of the electrode behave identically, rather than made by using one particle to replace all the particles \cite{Marquis2019}.
Therefore, the applied current $I_\text{app}$ needs to be shared by all active particles, and the input current $I_\text{in}$ for the diffusion-aware voltage source should be calculated as
\begin{align} \label{eq:inputcurrent}
	I_\text{in} = I_\text{app} / N_{\text{a}},
\end{align}
where $N_{\text{a}}$ is the number of active particles.

The simplified shell ECM in \cref{fig:fullECM}b captures all the physics of the single particle model (SPM)---lithium diffusion in an active particle and charge transfer overpotential---and thus can fully reproduce the SPM simulation results.
Moreover. it involves $R_\text{e}$ to model the electrolyte resistance within the separator.

In practice, the measured cell data do not distinguish between the NE and PE.
Thus, we further simplify the shell ECM in \cref{fig:fullECM}b to the one in \cref{fig:fullECM}c.
The resistance $R_0$ responsible for instantaneous voltage change includes the transport resistance in the whole cell including the electrolyte and electronically-conductive materials.
The diffusion-aware voltage source combines the diffusion effects of the PE and NE.
It is noteworthy that this ECM has only 2 parameters to calibrate: $R_0$ and $R_{\text{d},1}$. 

Finally, the terminal voltages for shell ECMs in \cref{fig:fullECM}b and c are calculated, respectively, as
\begin{subequations} \label{eq:terminalvolt}
\begin{align} 
	V_{\text{t}} & = \text{OCP}_\text{pe} - \text{OCP}_\text{ne} - I_\text{app} \qty(R_\text{ct,p} + R_\text{ct,n} + R_\text{e}), \\
	V_{\text{t}} & = \text{OCV} - I_\text{app} R_0.
\end{align}
\end{subequations}

\section{Comparison with physics-based model}
\label{sec:compspm}

This section aims to verify the developed diffusion-aware voltage source by comparing its simulation results with results of a physics-based model simulation.
Specifically, we first validate the equivalence between the diffusion-aware voltage source (\cref{fig:multivolts}) and the solid diffusion equation and verify its numerical implementation in \cref{sec:shellvoltsour}, and then in \cref{sec:verifyshellECM} we extend the verification to the shell ECM (\cref{fig:fullECM}b).
For simplicity, we choose the SPM and run it within PyBaMM~\cite{Sulzer2021} as the reference solution.
In the PyBaMM simulation, we pick the LG\,M50 cell with a parameter set characterized by Chen et al.~\cite{Chen2020}.
The parameters needed for simulations in this section can be found in \cref{tab:params}. 
\begin{table}
	\centering
	\footnotesize
	\caption{Parameters of a LG\,M50 cell~\cite{Chen2020} for simulations in \cref{sec:compspm}.}
	\label{tab:params}
	\renewcommand{\arraystretch}{1.3}
	\begin{threeparttable}
		\begin{tabular*}{\textwidth}{@{}l@{\extracolsep{\fill}}lcccl@{}} \toprule
			Parameter    & symbol       & unit       & \multicolumn{2}{c}{value} \\
			\midrule
			&        &        &  positive electrode & negative electrode \\
			\cmidrule (ll){4-5}
			particle radius   & $a$ & \si{\micro\meter} & 5.22 & 5.86  \\
			maximum lithium concentration & $c_{\text{s,max}}$ & \si{\mole\per\cubic\meter} & 63104 & 33133  \\
			initial concentration & $c_\text{s,0}$ & \si{\mole\per\cubic\meter} & \num{17038} & \num{29866}  \\
			active material volume fraction & $\epsilon_{\text{a}}$ & - & 0.665 &  0.75  \\
			diffusivity & $D_\text{s}$ & \si{\square\meter\per\second} & \num{4E-15} & \num{3.3E-14}  \\
			charge-transfer reaction rate &  $m$ & \si{\ampere \meter^{2.5} \mole^{-1.5}} &  \num{3.42e-6} & \num{6.48e-7}  \\
			electrode volume & $V$ & \si{\cubic\meter} & \num{7.764E-6} & \num{8.75E-6}  \\
			open circuit potential & OCP & \si{\volt} & \cref{fig:multivolts}c & \cref{fig:multivolts}c  \\
			\cmidrule (ll){4-5}
			cell nominal capacity & $Q$ & \si{\ampere\hour}  & \multicolumn{2}{c}{\num{5}} \\
			coefficient for unit conversion & $k$ & \si{\volt\cubic\meter\per\coulomb} & \multicolumn{2}{c}{\num{1}} \\
			lithium-ion concentration in electrolyte & $c_\text{e}$ & \si{\mole\per\cubic\meter} & \multicolumn{2}{c}{\num{1000}} \\
			Faraday constant  & $F$ & \si{\coulomb\per\mole}   & \multicolumn{2}{c}{\num{96485}} \\
			gas constant    & $R$   & \si{\joule\per\kelvin\per\mole}          & \multicolumn{2}{c}{\num{8.31}} \\
			absolute temperature & $T$ & \si{\kelvin}     & \multicolumn{2}{c}{\num{298.15}}  \\
			\bottomrule
		\end{tabular*}
	\end{threeparttable}
\end{table}

\subsection{Diffusion-aware voltage source}
\label{sec:shellvoltsour}

\begin{figure}
	\centering
	\includegraphics{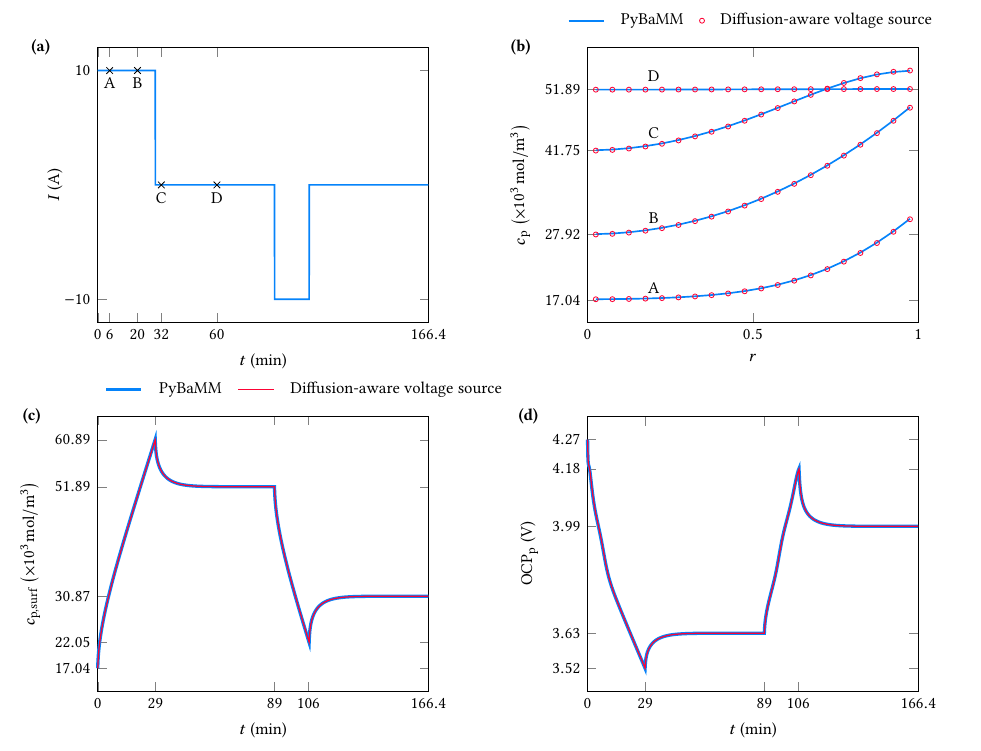}
	\caption{
		Verification of the diffusion-aware voltage source by comparing with PyBaMM simulation results.
		Subplot (a) shows the current profile of the test of discharge, rest, charge, and rest, and subplot (b) shows the concentration $c_\text{p}$ profiles of the positive electrode (PE) at 4 time instants A-D pinpointed in (a).
		Subplots (c) and (d) show the surface concentration $c_\text{p,surf}$ of the PE active particle and the corresponding PE open circuit potential (OCP\textsubscript{p}).
	}
	\label{fig:conc_grad}
\end{figure}

To check the ability of the diffusion-aware voltage source in resolving the lithium concentration, we simulate the response of a cell subjected to a sequence of constant-current discharge, rest, constant-current charge, and rest and monitor the concentration variation in the PE. 
The choice of the PE is because that it has lower diffusivity than the NE. 
The test protocol can be confirmed by the current profile in \cref{fig:conc_grad}a, where the sign of current is positive on discharge. 
From the PyBaMM simulation of SPM, we can retrieve the lithium concentration profiles (\cref{fig:conc_grad}b) at different time instants, the PE particle surface concentration evolution (\cref{fig:conc_grad}b), and the PE open circuit potential (\cref{fig:conc_grad}d) and use them to check the diffusion-aware voltage source response.

For a fair comparison, the same parameter values as in the SPM are used in the diffusion-aware voltage source (\cref{fig:multivolts}a).
Specifically, the following parameters are needed: particle radius $a$, diffusivity $D_\text{s}$, initial concentration $c_\text{s,0}$, and the number of particles $N_\text{a}$. 
The number of active particles is determined by two more parameters: 
\begin{align} \label{eq:particleno}
	N_\text{a} = \frac{\epsilon_\text{a} V}{ 4 \pi a^3 / 3},
\end{align}
where $V$ is the electrode volume and $\epsilon_\text{a}$ is the volume fraction of active material.

We pick four time instants (A-D) during the discharge and the following relaxation process and show the concentration spatial distribution for comparison in \cref{fig:conc_grad}b.
The surface concentration is calculated by \cref{eq:surfconcen} and presented in \cref{fig:conc_grad}c as a function of time.
Based on the surface concentration, the PE OCP is calculated according to the same function in the SPM.
The agreement in \cref{fig:conc_grad}b-d verifies the equivalence between the proposed circuit network and the discretized diffusion equation as well as the numeric implementation of the circuit network. 
We remark that this equivalence holds regardless of the applied currents and usage scenarios.

\subsection{Charge transfer resistance}

With the diffusion-aware voltage source verified, we further explore the possibility of reproducing the SPM using an ECM (\cref{fig:fullECM}b) incorporating the diffusion-aware voltage source. 
Besides the diffusion overpotential, another major contribution to terminal voltage loss in the SPM comes from the reaction overpotential.
The reaction overpotential in the SPM does not involve a timescale, and thus we simply use a resistor with varying resistance to reproduce the same effect.

To obtain this charge transfer resistance, we first calculate the reaction overpotential by reversing the Butler-Volmer equation \cite{Doyle1996} that describes the electrochemical reaction at the active particle surface.
According to Butler-Volmer equation, the interfacial current density $j$ is expressed as~\cite{Marquis2019}
\begin{align} \label{eq:butlervolmer}
	j = 2 j_0 \sinh \qty(\frac{F \eta}{2RT}),
\end{align}
where~$j_0$ is the exchange current density and~$\eta$ is the reaction overpotential.
In the SPM, the interfacial current density is determined from the applied current $I_\text{app}$ via
\begin{align}
	j = \frac{I_\text{app}}{A} = \frac{I_\text{app}}{3 \epsilon V / R} = \frac{I_\text{app}}{S_\text{a} \epsilon V},
\end{align}
where $A$ is the total surface area of active materials, $S_\text{a} = 3 / R$ denotes the particle specific area (surface area to volume ratio), $\epsilon$ is the active material volume fraction, and $V$ represents the electrode volume.
The inverse Butler-Volmer relation is then formulated to express the reaction overpotential $\eta$, and the charge transfer resistance is then calculated as
\begin{align} \label{eq:rct}
	R_\text{ct} = \frac{\eta}{I_\text{app}} = \frac{2RT}{FI_\text{app}} \text{arcsinh} \qty(\frac{j}{2j_0}) =
	\frac{2RT}{FI_\text{app}} \text{arcsinh} \qty(\frac{I_\text{app}}{2 j_0 S_\text{a} \epsilon V}).
\end{align}
Here, the exchange current density~$j_0$ is written as~\cite{Marquis2019}
\begin{align} \label{eq:exchangej}
	j_0 = m \sqrt{c_{\text{s,surf}} \qty(c_{\text{s,max}} - c_{\text{s,surf}}) c_{\text{e}}},
\end{align}
where~$m$\,$\qty(\si{\ampere\per\square\meter \cdot (\cubic\meter\per\mole)^{1.5}})$ is a rate constant of the charge transfer reaction, $c_{\text{s,max}}$ is the maximum lithium concentration in the active particle, $c_{\text{s,surf}}$ is the surface concentration of the active particle, and $c_{\text{e}}$ is the lithium-ion concentration in the electrolyte.
In the SPM, the lithium-ion concentration in electrolyte is simplified as a constant, and thus the exchange current density $j_0 \qty(c_{\text{s,surf}})$ is only a function of particle surface concentration $c_{\text{s,surf}}$. 
Finally, we remark that $R_\text{ct}$ in \cref{eq:rct} expresses the equivalent resistance of the charge transfer resistors of all particles within each electrode arranged in parallel because the applied current is assumed to be shared by all particles equally. 

The inverse hyperbolic sine function $\text{arcsinh} \qty(x)$ approaches $x$ as $x$ tends to $0$, i.e.,
\begin{align}
	\lim_{x \to 0} \frac{\text{arcsinh} (x)}{x} = 1.
\end{align}
Hence, at low currents, we approximate $\text{arcsinh} \approx x$, and \cref{eq:rct} is further simplified as~\cite{Geng2021}
\begin{align} \label{eq:rct-app}
	R_\text{ct} \approx \frac{RT}{F j_0 S_\text{a} \epsilon V}. 
\end{align}
In this case, the charge transfer resistance is solely determined by the particle surface concentration (the electrolyte concentration $c_{\text{e}}$ in \cref{eq:exchangej} is assumed constant in the SPM; see \cref{tab:params}). 
The particle surface concentration is time-varying, so is the charge transfer resistance.

\begin{figure}
	\centering
	\includegraphics{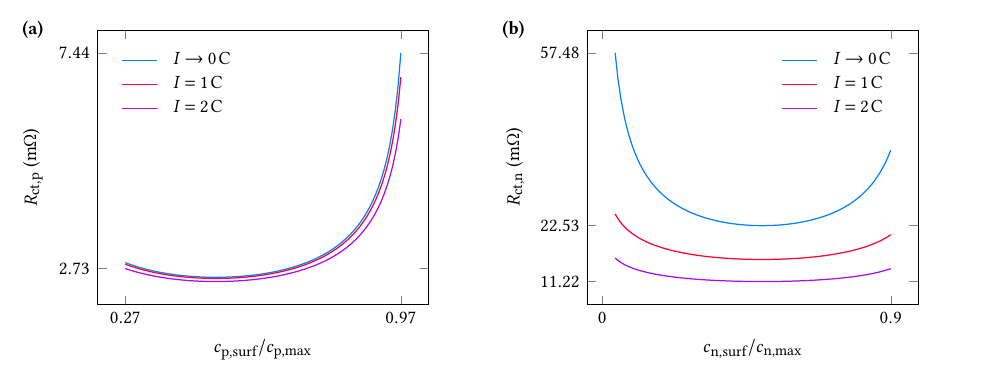}
	\caption{
		Charge transfer resistance in the positive ($R_\text{ct,p}$) and negative ($R_\text{ct,n}$) electrodes as function of surface concentration at different C-rates.
	}
	\label{fig:rct}
\end{figure}

To check the effect of applied current magnitude, we calculate the charge transfer resistance at different C-rates as shown in \cref{fig:rct}.
In general, the resistance of the NE is an order of magnitude larger than that of the PE (caused by the reaction constant $m$ in \cref{eq:exchangej}), and the effect of the applied current is more significant for the NE.
For both electrodes, the approximation at small currents overestimates the resistance.

\subsection{Shell ECM}
\label{sec:verifyshellECM}

\begin{figure}
	\centering
	\includegraphics{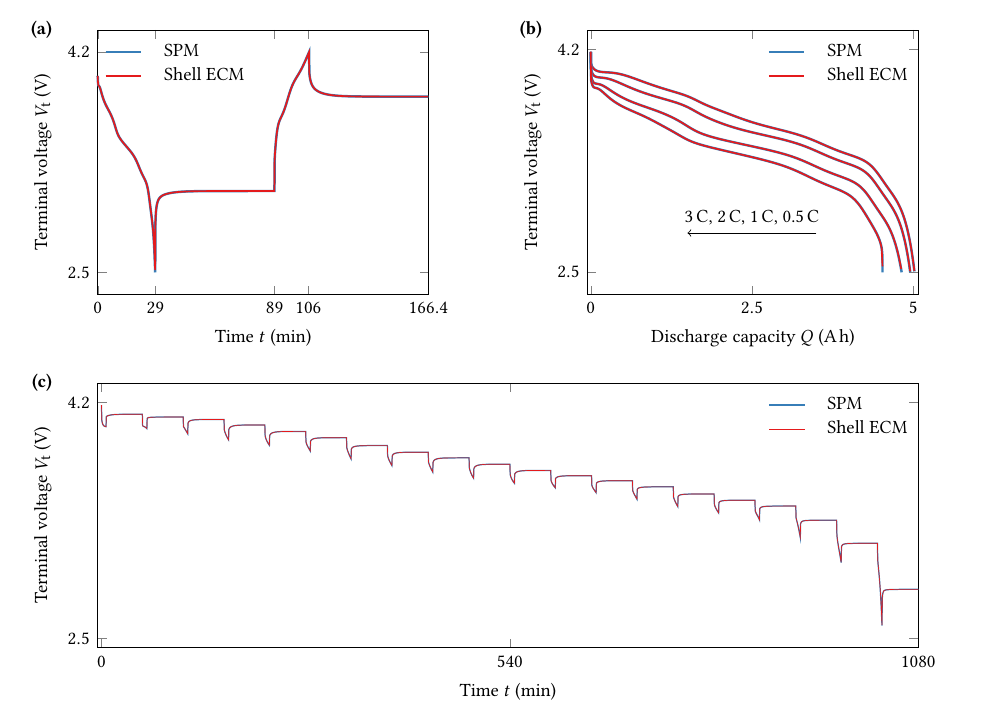}
	\caption{
		Comparison between shell ECM results and results from SPM simulation in PyBaMM in terms of terminal voltage for three test protocols.
	}
	\label{fig:comp-spm}
\end{figure}

We further compare the shell ECM (\cref{fig:fullECM}b) against the SPM. 
For the shell ECM, the electrolyte resistance $R_\text{e}$ is set to be null, and charge transfer resistances $R_\text{ct,n}$ and $R_\text{ct,p}$ are determined through \cref{eq:rct}. 
Three typical test profiles are considered: a discharge-rest-charge-rest cycle (same as in previous section), Galvanostatic Intermittent Titration Technique (GITT), and constant-current discharges at varying current rates.
\cref{fig:comp-spm} shows the terminal voltage obtained from running the shell ECM and the PyBaMM SPM simulation with the same parameters. 
It is shown that the shell ECM can completely reproduce the SPM simulation results.

\section{Experimental validation}
\label{sec:expval}

This section aims to validate the proposed shell ECM by comparison with experimental measurements of LG\,M50T cells.
In real-world applications, the terminal voltage is often measured for the whole cell without differentiating the contributions from the two electrodes. 
Thus, we choose the simplified shell ECM involving a single diffusion-aware voltage source as shown in \cref{fig:fullECM}c.

For parameterization and validation, GITT and constant-current discharge tests were performed respectively on the LG\,M50T cells. 
Before these tests, the cells were fully charged through the constant-current/constant-voltage protocol at a charge rate of 0.3\,C to a maximum voltage of 4.2\,V until a cut-off current of C/100, followed by a two-hour rest. 
The GITT tests consists of 25 repeating segments of a discharge pulse at a 1\,C rate for 144 seconds followed by a one-hour rest. 
The cell nominal capacity is \SI{5}{\ampere\hour}, and thus each discharge pulse consumes a capacity of \SI{0.2}{\ampere\hour}. 
Constant-current discharge tests were conducted at rates of 0.4\,C and 2\,C until the lower voltage limit of 2.5 V was reached. 
All tests were performed at a temperature of \SI{25}{\degreeCelsius} in a thermal chamber.
Full details of the experimental procedures, apparatus, and thermal management can be found in \ref{app:exp}.

The shell ECM is first parameterized using the GITT data and then used to predict the cell performance under constant-current discharge conditions. 
The model predictions are compared with the measurements. 

\subsection{Parameterization}
\label{sec:parameterization}

\begin{figure}
	\centering
	\includegraphics{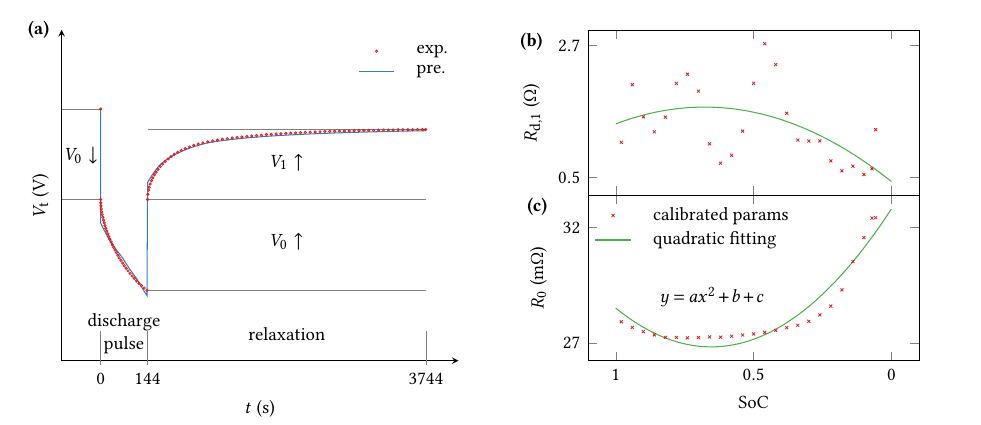}
	\caption{
		The two shell-ECM parameters are determined for each discharge-rest segment. 
		(a) A typical discharge pulse at a 1\,C rate followed by a rest where the two parameters are assumed constant.
		(b) Calibrated parameters for 25 discharge-rest segments and the second-order polynomial fitting of these parameters as function of the state of charge (SoC) for model prediction in the following constant-current discharges. 
		The SoC of each point takes the mean value of the SoC range that the corresponding discharge-rest segment goes through. 
	}
	\label{fig:layer_method}
\end{figure}

For the shell ECM in \cref{fig:fullECM}c, there are only two parameters to be calibrated: the ohmic resistance $R_0$ and the diffusion resistance $R_\text{d,1}$ between the first and the second layers (\cref{fig:multivolts}).
These two parameters are determined from the GITT data on a pulse-by-pulse basis as sketched in \cref{fig:layer_method}a. 
For each discharge pulse followed by a rest, the two parameters are assumed constant and are identified separately. 
The ohmic resistance $R_0$ is determined from the instantaneous drop and increase of the terminal voltage $V_0$ divided by the applied current, and the average of the two $R_0$ values is recorded. 
The diffusion resistance $R_\text{d,1}$ is fitted by minimizing the difference between the experimental measurements and model prediction of the terminal voltage (i.e., least-squares method). 
The least-squares fitting is only applied to the relaxation section in \cref{fig:layer_method}a where the ohmic effect vanishes. 

The calibrated parameters for each discharge-rest segment are then presented as discrete points in \cref{fig:layer_method}b and c.
Each discharge pulse undergoes \SI{144}{\second} at a 1\,C rate and corresponds to $4\%$ of the cell nominal capacity.
Thus, there are in total 25 discharge pulses and corresponding parameter sets. 
The cell average SoC is calculated by Coulomb counting and its variation during each discharge pulse is recorded.
The mean value of the two SoC limits of each pulse is then used to indicate the SoC level associated with the calibrated parameters.

\begin{figure}
	\centering
	\includegraphics{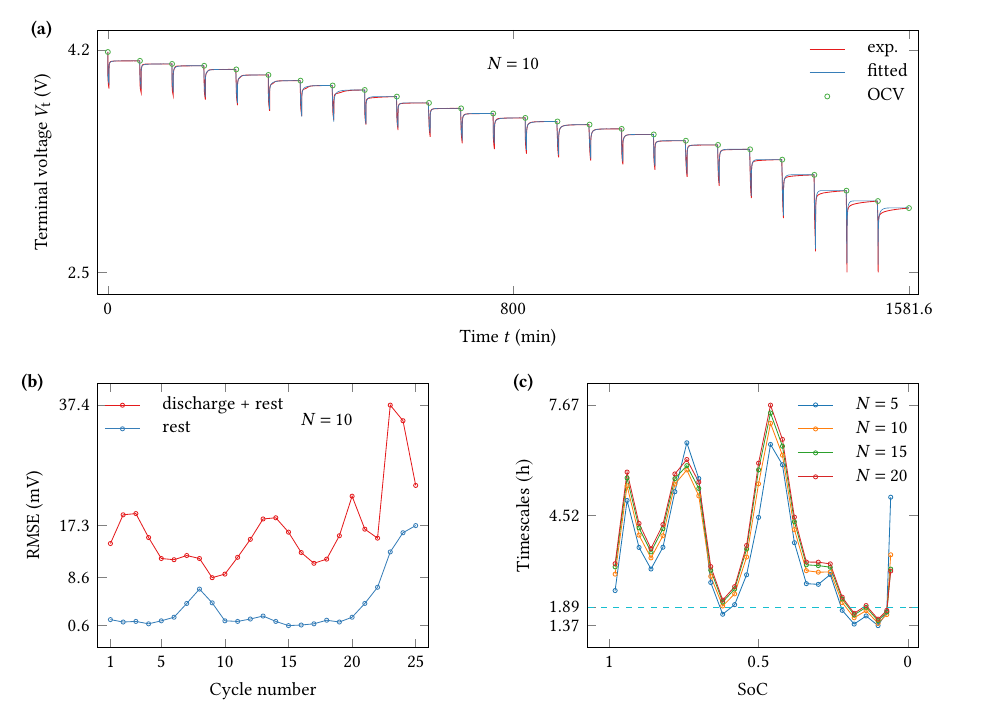}
	\caption{
		(a) Best-fitting curve against measured data with calculated $R_0$ and fitted $R_\text{d,1}$. 
		(b) Root mean square errors of the best-fitting curve for each discharge-rest segment. 
		(c) Diffusion timescales in \cref{eq:timescale-fit} derived from the fitted diffusion resistances.
		For the timescale estimation, we explored the effect of layer number; otherwise, $10$ layers are used elsewhere.
	}
	\label{fig:exp-gitt-volt}
\end{figure}

With the two parameters identified for each segment, the best-fitting voltage curve is shown in comparison with the experimental curve in \cref{fig:exp-gitt-volt}a. 
Also, the equilibrium points after relaxation are taken to serve as the OCV. 
To evaluate the calibration, we plot the root mean square error (RMSE) for each discharge-rest segment in \cref{fig:exp-gitt-volt}b:
\begin{align}
	\text{RMSE} = \sqrt{\frac{1}{J} \sum^{J}_{j=1} {\qty(V_{\text{exp},j} - V_{\text{pre},j})^2}},
\end{align}
where $j$ represents the $j$-th point and $J$ is the number of data points. 
Here we present two RMSE curves: one takes data of the whole discharge-rest segment, indicating the overall error of calibrating $R_0$ and fitting $R_\text{d,1}$; the other only uses the data of the rest section, indicating the least-squares fitting error. 
Compared to the overall error, the fitting error of $R_\text{d,1}$ is relatively small, implying that the relaxation process through diffusion is well modeled. 
Another observation in \cref{fig:exp-gitt-volt}b is that the fitting errors in the low SoC area (pulses 23--25) are higher than those for other pulses. 
This observation suggests that the poor estimation accuracy is not likely to be improved by considering the solid diffusion and resolving surface concentration for the LG\,M50T cells, different from the conclusion by Ouyang et al.~\cite{Ouyang2014}.

According to \cref{eq:timescale}, the timescale of solid diffusion is related to the diffusion resistance $R_\text{d,1}$.
Here, the diffusion resistance $R_\text{d,1}$ is fitted from experimental data and reflects the collective effect of diffusion in both electrodes.
In general, the NE has diffusivity one order of magnitude higher than that of the PE \cite{Chen2020}, that is, the diffusion in the PE is the limiting process. 
We thus assume that the fitted resistance mainly indicates the PE diffusion timescale, and the timescale is then estimated as
\begin{align} \label{eq:timescale-fit}
	\tau_\text{fit} = \frac{4 \pi a^3 F c_\text{s,max} N_\text{a}}{kN} R_\text{d,1},
\end{align}
where $N_\text{a}$ is the particle number calculated by \cref{eq:particleno} and $N$ is the number of layers of the diffusion-aware voltage source (\cref{fig:fullECM}c). 
The derivation of \cref{eq:timescale-fit} can be found in \ref{app:fittimescale}.
The fitted timescale $\tau_\text{fit}$ would not change too much if the parameter set of NE is used, in view of the consistency between both electrodes in terms of capacity ($\sim c_\text{s,max} a^3 N_\text{a}$). 

The fitted timescales at different SoC levels basically remain at the same magnitude and generally higher than the value of \SI{1.89}{\hour} calculated from PE particle radius and diffusivity of a LG\,M50 cell \cite{Chen2020}. 
We remark that the difference caused by number of layers is negligible, confirming that the layer number does not change the physics but affects the solution accuracy.

\begin{figure}
	\centering
	\includegraphics{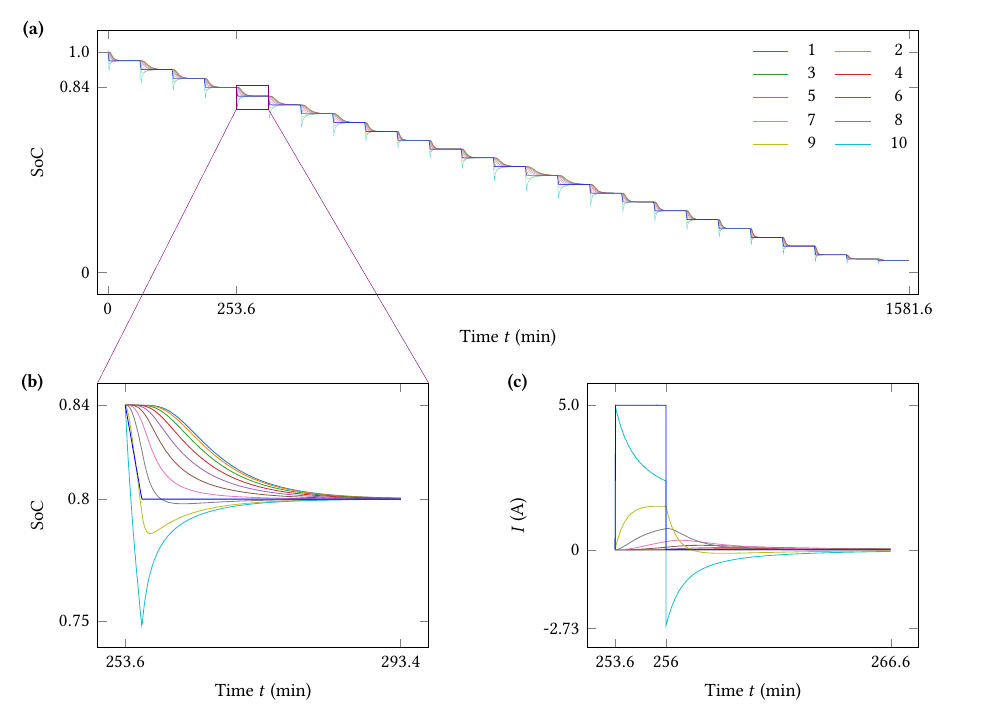}
	\caption{
		Evolution of the SoC of each layer and the average SoC (a), and the fifth discharge-rest segment is magnified in (b) for clarity.
		The corresponding current outflow from each layer is plotted in (c), where the sign of current is opposite to that in \cref{fig:multivolts}. 
		Note that the diffusion-aware voltage source, as well as the associated particle, represents the collective effects of both electrodes. 
	}
	\label{fig:exp-gitt-soc}
\end{figure}

\cref{fig:exp-gitt-soc}a shows the evolution of the SoC of each layer as well as the cell average SoC at the fitted diffusion resistance.
The average SoC decreases during the discharge pulse and then remains unchanged during the rest period; this pattern is repeated from segment to segment.
A typical segment (the fifth) is then demonstrated in \cref{fig:exp-gitt-soc}b.
The outermost layer (10th) is characterized by the fastest decrease of SoC in the beginning of the discharge pulse, while the innermost layer does not lose any lithium, forming a concentration gradient inside the particle.
After the current is turned off, the concentration in the outermost layer immediately starts to increase while lithium keeps flowing from the inner layers to the outer layers towards an equilibrium.
The derivative of the concentration with respect to time is proportional to the current flowing out from each layer in \cref{fig:exp-gitt-soc}c.
The current $I_{10}$ flowing out the 10th layer contributes most to the discharge pulse, followed by the 9th layer.
The sudden drop of $I_{10}$ at \SI{256}{\minute} from a positive (discharge) value to a negative (charge) one explains the sharp reversing of SoC\textsubscript{10} in \cref{fig:exp-gitt-soc}b.
The difference between SoC\textsubscript{10} and the average SoC is responsible for the diffusion overpotential, because that SoC\textsubscript{10} is close to the surface SoC. 

\subsection{Prediction}
\begin{figure}
	\centering
	\includegraphics{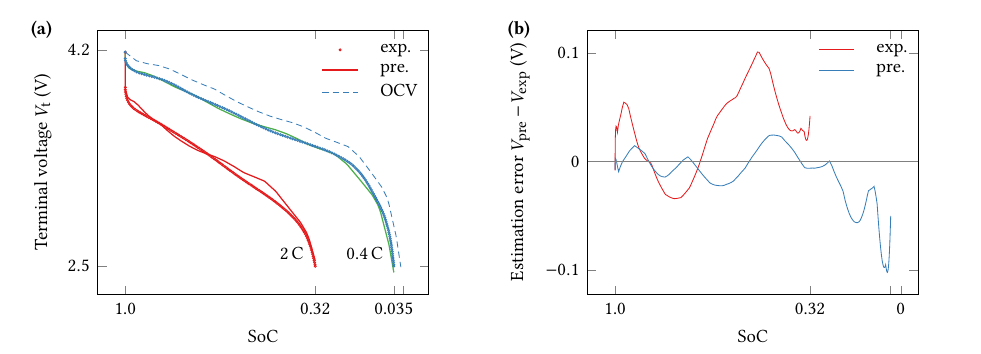}
	\caption{
		Comparison between the shell ECM prediction and experimental data.
	}
	\label{fig:exp-cc-pred}
\end{figure}

In this section, we use the calibrated two parameters ($R_0$ and $R_\text{d,1}$) to predict the cell performance in constant-current discharge tests and compare with the experimental data.
The two datasets in \cref{fig:layer_method}b and c, expressing the dependency of $R_\text{d,1}$ and $R_0$ on the cell SoC, are fitted by quadratic functions using the least-squares method. 
The quadratic fitting in \cref{fig:layer_method}b smoothes out the diffusion ``spikes'' (also reported in Fig. 6 of \cite{ORegan2022}) in order to match with the smooth voltage curves in \cref{fig:exp-cc-pred}a. 
The two best fits (green solid lines) are then used in the shell ECM.
Since these parameters are obtained from GITT data at a 1\,C discharge rate, we pick two other rates, 0.4\,C and 2\,C, for the constant-current discharges.

The model prediction are presented in \cref{fig:exp-cc-pred}a, in comparison with the experimental measurements.
The estimation error $V_\text{pre} - V_\text{exp}$, the difference between the predicted terminal voltage and measured voltage, is found to be restricted to the interval $+/-$\SI{0.1}{\volt}, as shown in \cref{fig:exp-cc-pred}b.
The prediction captures the trend of the terminal voltage variation at a low and a high current rate, suggesting that the two overpotentials caused by ohmic resistance and diffusion are the major contributions to the terminal voltage variation.
The predicted voltage curves are not as smooth as the experimentally measured curves at the two discharge rates; this is however expected and does not affect the results and interpretation. 
The voltage curves are based on the dashed OCV curve in \cref{fig:exp-cc-pred}a, which is linearly interpolated from the 25 discrete OCV data points in the GITT data in \cref{fig:exp-gitt-volt}a. 
Further smoothing of these OCV data points would lead to a smoother OCV curve and smoother voltage curves. 
Finally, we remark that the rest time (one hour) between two consecutive discharge pulses is relatively short and thus the obtained OCV points especially in the low SoC region may not correspond to the cell equilibrium states, lowering the prediction accuracy. 

\section{Implementation into battery management systems}
\label{sec:imple}

In this section, we implement the shell ECM into a battery management system (BMS) and evaluate its performance in comparison with a traditional ECM. 
Specifically, the implementation is achieved within the closed-loop state estimation BMS of WAE Technologies (WAE). 
Traditional ECMs are implemented in present-day onboard battery management algorithms to maintain low computational cost, and they have been proven to work well for model-based power limits estimation~\cite{Plett2004}. 
To evaluate the viability of the shell ECM for practical onboard usage, we test the shell ECM within the WAE power limits algorithm in a model-in-the-loop testing environment and compare the computational time incurred by the shell ECM with that by the traditional ECM. 

\begin{figure}
	\centering
	\includegraphics{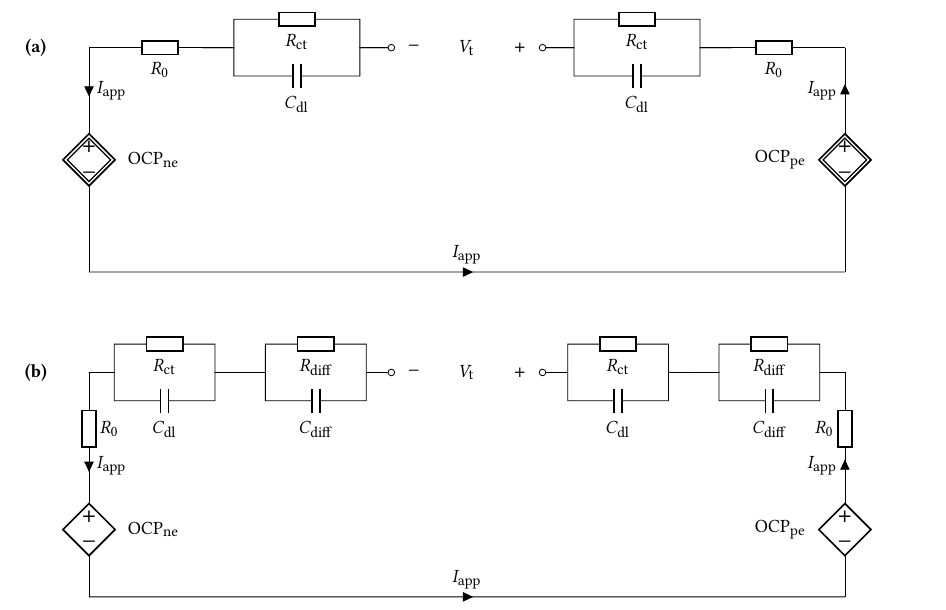}
	\caption{
		Configuration of (a) the shell ECM and (b) traditional ECM implemented into WAE Technologies battery management systems. 
	}
	\label{fig:shellvsold}
\end{figure}

The shell ECM and traditional ECM we implemented are sketched in \cref{fig:shellvsold}. 
The shell ECM here differs from the SPM-equivalent one (\cref{fig:fullECM}b) in that it uses an RC pair (charge transfer resistor $R_\text{ct}$ and double layer capacitor $C_\text{dl}$) to account for the fast dynamics and an ohmic resistor $R_0$ for the overpotential contributed by electronic current flow in the solid and ionic current flow in the electrolyte. 
The diffusion-aware voltage source for each electrode has 5 layers and thus 5 internal states of concentration to track. 
For a fair comparison, the traditional ECM also has a symmetrical structure for the two electrodes~\cite{Zhang2022}. 
Since the traditional voltage source cannot resolve concentration gradient, the diffusion overpotential is considered by an additional RC pair ($R_\text{diff}$ and $C_\text{diff}$). 
The rest is the same as the shell ECM---an RC pair for fast dynamics and an ohmic resistor for instantaneous voltage change. 

The two models are parameterized against half-cell pulse data collected from WAE's physics-based model of the LGM50 cell. 
The procedures and data for parameterization are the same for both the shell and traditional ECMs.
The parameterized shell ECM and traditional ECM are then evaluated in an industrial drive-cycle test. 
The drive-cycle test data are also generated from WAE's physics-based model, featuring discharge and regeneration (charge) events (\cref{fig:wae_pred}a).
In the following, we compare the shell ECM with the traditional ECM in terms of the computational cost and prediction accuracy. 

With reference to \cref{fig:shellvsold}b, the traditional ECM totalled 10 parameters: 8 for the 4 RC pairs and 2 for the 2 ohmic resistors. 
It has 5 states to compute: one state for each RC pair and an extra state for the SoC. 
According to \cref{fig:shellvsold}a, the shell model totalled 8 parameters---each diffusion-aware voltage source has one independent diffusion resistance. 
However, it has 12 states for computation: 2 states for 2 RC pairs, and 10 states for the concentrations in the 10 particle layers (5 for each electrode). 
Here the SOC of each electrode is calculated based on the volume average of the concentrations in the total 5 layers; unlike the traditional ECM, no state is needed for the average SoC. 

\begin{figure}
	\centering
	\includegraphics{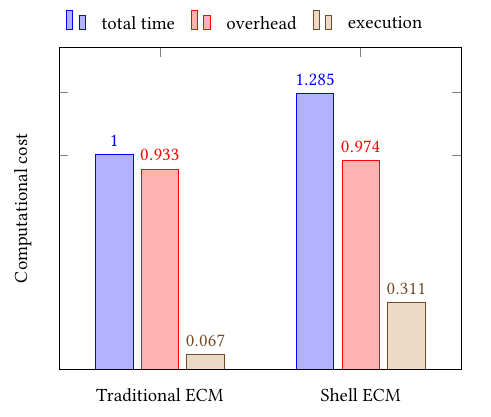}
	\caption{
		Computational cost decomposition and comparison between the shell ECM and the traditional ECM. 
		The total computational time is the sum of the model execution time and the overheads by parameter look-up. 
		All the costs are normalized by the total computational time of the traditional ECM. 
	}
	\label{fig:cost}
\end{figure}

\begin{figure}
	\centering
	\includegraphics{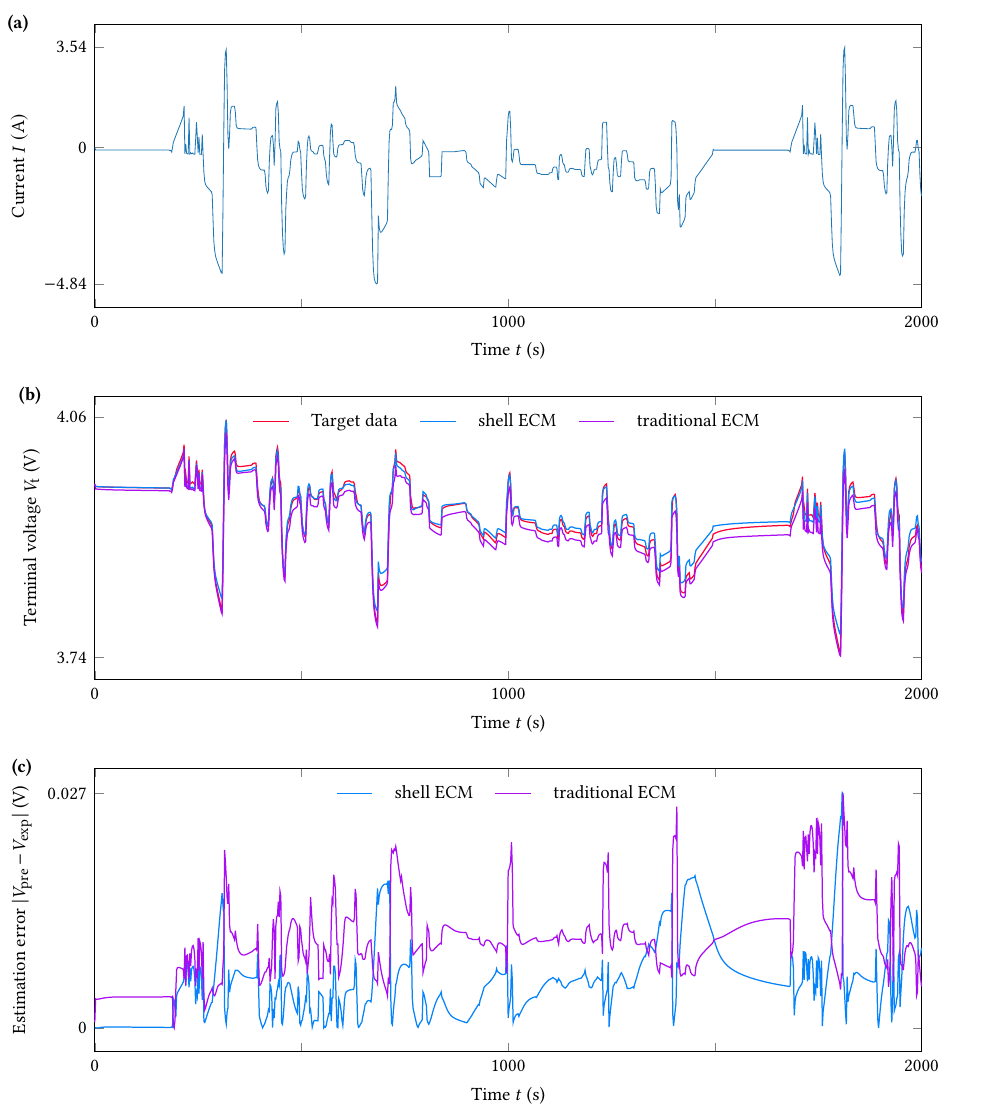}
	\caption{
		Comparison of the shell ECM with the traditional ECM in prediction of the terminal voltage (b) of a drive-cycle test with the current profile in subplot (a).
		The estimation error in subplot (c) is demonstrated as the absolute value of the difference between the prediction and target data, which is generated by WAE's physics-based model.
	}
	\label{fig:wae_pred}
\end{figure}

\cref{fig:cost} shows the computational cost of the two models by profiling analysis. 
All the computational costs are normalized by the total computational time of the traditional ECM. 
For both models, the total computational cost is decomposed into the model execution time and parameter lookup overheads. 
The model execution time of the shell ECM is 4.6 times longer than that of the traditional ECM (0.311 vs 0.067).
This is well expected as the shell ECM is equivalent to the discretized diffusion equation and has 12 states to compute. 
However, in both cases the model execution time is relatively insignificant compared to the overheads by parameter lookup: the execution time is 6.7\% and 24.2\% of the total computational time for the traditional ECM and shell ECM, respectively. 
In total, the cost of the shell ECM is 1.3 times the cost of the traditional ECM, when both implemented within a well-developed and computationally efficient BMS. 
This small increase of overall computational cost is insignificant, in the context of a BMS's total computational overheads. 
Moreover, the cost of the shell ECM can be further reduced after numerical techniques and layer number are optimized. 

The shell ECM shows advantages in terms of prediction accuracy. 
\cref{fig:wae_pred} shows an overall higher prediction accuracy of the shell ECM compared to the traditional ECM in predicting the terminal voltage of a drive-cycle test.
Subplot (a) shows the current evolution, subplot (b) shows the predicted terminal voltages by the two models in comparison with the generated data (cited as target data hereafter) by WAE's physics-based model, and subplot (c) shows the difference between the prediction and target data. 
It can be seen that large errors of the traditional ECM tend to occur following high current pulses. 
To confirm this, we further investigate the two models subjected to a charge pulse at a high current rate (2\,C) followed by a rest. 

\begin{figure}
	\centering
	\includegraphics{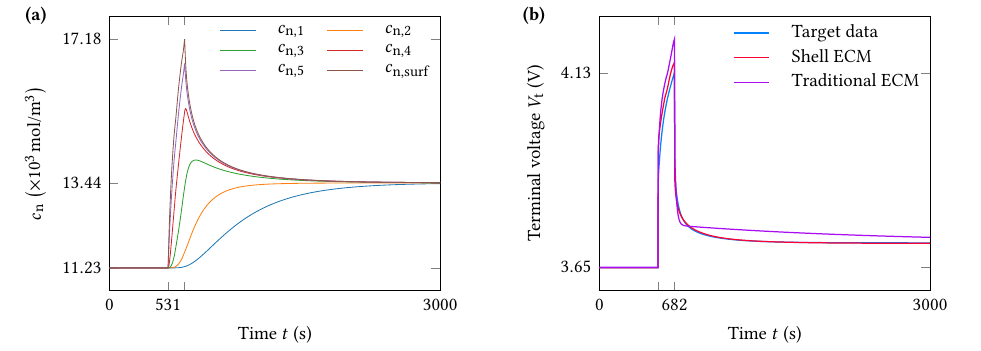}
	\caption{
		A test case of 2C charge pulse followed by a rest: (a) the lithium concentration $c_\text{n}$ evolution in the 5 layers of negative electrode particle recorded by WAE's battery management system and (b) comparison between the two models in predicting the terminal voltage. 
	}
	\label{fig:wae_pred_cc}
\end{figure}

\cref{fig:wae_pred_cc}b shows that the shell ECM well captures the terminal voltage during relaxation, while a noticeable discrepancy is observed between the traditional ECM prediction and the target response. 
The traditional ECM lacks physics; in practice, its parameters are often set to be dependent on the terminal current to maintain a desired level of accuracy when the electrode particle is highly polarized. 
The current dependency however leads to significant inaccuracy especially during cell relaxation following a charge or discharge: as the current drops close to zero, the model parameters take the values for low current rates, while the large concentration gradient persists. 
Although numerous attempts have been made to improve the traditional ECM, the above-mentioned issue remains. 
The shell ECM avoids this issue because it models and tracks the internal concentration states as shown in \cref{fig:wae_pred_cc}a, with robustness to dynamic current profiles. 
Moreover, the parameter of the shell ECM---diffusion resistance---has physical meaning and reduced dependency on external current. 

By modeling internal states, the shell ECM enables the application of advanced power limit strategies~\cite{Zheng2018} to improve cell performance and mitigate cell degradation. 
For example, the surface concentration and diffusion resistance can be used to control power limits based on the anode potential and visibility of internal concentration gradients, as opposed to the cell voltage alone, thus managing lithium plating risk and enabling faster charging~\cite{Xavier2021}. 

\section{Conclusion}

We proposed a circuit network (\cref{fig:multivolts}) to describe the lithium diffusion in an active particle, and the proposed circuit network is theoretically and numerically proven equivalent to the discretized diffusion equation based on finite volume method.
We define the whole circuit network as a new high-level circuit element, called diffusion-aware voltage source and symbolized by a double-wall diamond (\cref{fig:fullECM}).
The diffusion-aware voltage source gives the electrode potential as a function of particle surface concentration, acting as an alternative to the combination of a traditional voltage source and RC pairs in standard ECMs.

Inside the proposed circuit network, the linear dependence of the voltage of each layer on the local concentration/SoC ensures that the concentration gradient is the driving force. 
The real OCP relation depending on the surface concentration is used as the output of the diffusion-aware voltage source. 
The diffusion-aware voltage source has been verified by comparison against SPM simulations within PyBaMM, and the simplest shell ECM has been validated by comparison against experimental data. 

The value of the work lies in the circuit analogy to the spatially discretized diffusion equation and boundary conditions. 
This analogy transforms the continuous description of physics-based models into well-established knowledge/behavior of circuit elements, thereby reducing the implementation complexity for practical on-board usage. 
Compared to traditional ECMs, the shell ECM containing diffusion-aware voltage sources demands a comparable computational cost, but demonstrates superiority in terms of prediction accuracy and robustness to dynamic current profiles, when both models are tested in a battery management system. 
By tracking the internal concentration states with a high fidelity, the shell ECM shows great promise in the application of advanced power limit strategies to mitigate cell degradation and enable fast charge.

\section*{Acknowledgements}
The research leading to these results has received funding from the Innovate UK through the WIZer Batteries project (grant number 104427) and EPSRC Faraday Institution Multi-Scale Modelling project (EP/S003053/1, grant number FIRG025).

\section*{Data availability}
All relevant data leading to the findings of this study, including parameter values and source codes for model implementation, are openly available in the GitHub repository at \url{https://github.com/mzzhuo/DAVS}.

\appendix

\section{Numerical method}
\label{app:num}

This appendix provides numerical procedures to solve the diffusion-aware voltage source as shown in \cref{fig:multivolts}. 
The diffusion-aware voltage source serves as spatial discretization of the diffusion equation and converts a partial differential equation into 
a system of ordinary differential equations. 
Numerical solutions to ordinary differential equations have been well addressed in textbooks, and open-source solvers are readily available. 
Depending on desired accuracy and implementation complexity, different methods can be used to solve the system of internal state evolution equations~(\ref{eq:connection}). 

Here we present the time stepping schemes of the forward and backward Euler methods. 
The temporal discretization of \cref{eq:connection} for layer $n$ using the forward and backward schemes can be expressed, respectively, as
\begin{align}
	\frac{c_n^{i+1} - c_n^{i}}{\Delta t} &= \frac{1}{F\Omega_n} I_n^i, \label{eq:forward}\\
	\frac{c_n^{i+1} - c_n^{i}}{\Delta t} &= \frac{1}{F\Omega_n} I_n^{i+1}, \label{eq:backward}
\end{align}
where $i$ denotes the current $i$-th time step. 
The forward Euler method~(\ref{eq:forward}) is an explicit scheme and thus very easy to implement, but it suffers from numerical instability. 
The stability condition of the forward scheme for states of the diffusion-aware voltage source reads
\begin{align}
	\Delta t \le \frac{1}{2} \frac{b^2}{D_\text{s}}, 
\end{align}
where $\Delta t$ is the time-step length, $b$ is the thickness of each layer, and $D_\text{s}$ is the diffusivity. 
The backward Euler method~(\ref{eq:backward}) is more difficult to implement but is free from stability issues. 
In our case, the diffusivity $D_\text{s}$ is constant and thus \cref{eq:backward} is a linear system of equations, requiring comparable computational cost compared to the explicit scheme. 
Both the forward and backward Euler methods are of first-order accuracy; for higher-order accuracy, we can resort to improved Euler method or Runge-Kutta methods. 

\section{Experimental method}
\label{app:exp}

This appendix details the experimental setup for the measured data presented in \cref{sec:expval}. 
Experiments were performed on LG\,M50 (LG GBM50T2170) cylindrical lithium-ion battery cells. 
All electrochemical data was recorded by a Biologic BCS-815 battery cycler with the accompanying BT-Lab software. 
The temperature of the cell was recorded using K-type thermocouples adhered to the cell surface using Kapton tape, approximately halfway along the axial direction. 
These thermocouples were connected to the built-in thermocouple readers of the BCS-815 battery cycler, with temperature data recorded alongside the electrochemical data.

Experiments at the low discharge rate (0.4C) were performed with the cell housed inside a binder thermal chamber (KB 23 cooling incubator), set to maintain a stable air temperature of \SI{25}{\degreeCelsius}. 
The fan speed of the thermal chamber was set to 100\%, and the cell was placed in a regime cooled by forced air convection. 
Electrical connections to the cell were made via a spring-loaded cell holder, which provided a 4-point connection via banana plugs. 
Connection resistances were not compensated in subsequent tests.

Experiments at higher discharge rates (1C and 2C) were performed with the cell in a conductive cooling regime in order to limit the temperature rise. 
The cell was in direct thermal contact with aluminium blocks, which had been machined to fit around the cylindrical surface of the cell. 
Thermal interface material was used to improve heat transfer between the cell and the blocks. 
The aluminium blocks were held at a constant temperature of 25°C using a bespoke temperature controller. 
The temperature controller used Peltier elements to heat/cool the blocks based on a PID control, with K-type thermocouples providing feedback. 
The design of these controllers is described in more details in our previous work~\cite{Kirkaldy2022}. 
Electrical connections to the cell were made via spot-welded nickel strips which were clamped between brass blocks, and the brass blocks were connected to the battery cycler through banana plugs.

\section{Diffusion timescale estimation}
\label{app:fittimescale}

This appendix introduces the derivation of \cref{eq:timescale-fit}: how to estimate the diffusion timescale (including diffusivity) from the fitted diffusion resistance? 
In \cref{sec:parameterization}, the parameter $R_\text{d,1}$ of the diffusion-aware voltage source (\cref{fig:fullECM}c) is fitted against experimental GITT data (\cref{fig:layer_method}).
Note that the single diffusion-aware voltage source represents the collective diffusion effect of both electrodes, and at this stage we have information of neither electrode in terms of the size and number of active particles. 
We thus take the local SoC $z_n$ as the internal state for layer $n$ that is calculated as
\begin{align} \label{eq:diff_zn}
	\frac{\dd{z_n}}{\dd{t}} = \frac{I_n}{Q_{n}},
\end{align}
where $Q_{n}$ is the volume-weighted fraction of the cell nominal capacity $Q$. 
We re-define the controlled voltage as
\begin{align} \label{eq:controlledvolt2}
	V_{n} = kz_{n},
\end{align}
and substitute it into \cref{eq:circuitgov}, leading to
\begin{align} \label{eq:shelleq2}
	I_n & = \frac{z_{n+1} - z_{n}}{R_{\text{d},n} / k} - \frac{z_{n} - c_{z-1}}{R_{\text{d},n-1} / k}.
\end{align}
\cref{eq:diff_zn,eq:controlledvolt2,eq:shelleq2} are complete to fit the diffusion resistance $R_\text{d,1}$ with a given applied current $I_\text{app}$ as the input. 

To estimate the diffusion timescale, we need to relate the fitted diffusion resistance in \cref{eq:shelleq2} to the diffusivity in the discretized diffusion equation~(\ref{eq:discretized}). 
The cell nominal capacity is assumed to be equal to the capacity of either electrode and estimated as 
\begin{align}
	Q = F c_\text{s,max} \Omega N_\text{a},
\end{align}
where $\Omega$ is the active particle volume. 
This relation holds for either of the two electrodes. 
Accordingly, the nominal capacity for layer $n$ is $Q_n = F c_\text{s,max} \Omega_n N_\text{a}$, and \cref{eq:diff_zn} can thus be reformulated as
\begin{align} \label{eq:diff_zn2}
	I_n = F c_\text{s,max} N_\text{a} \frac{\dd{z_n}}{\dd{t}} \Omega_n.
\end{align}
Dividing both sides of \cref{eq:discretized} by $c_\text{s,max}$ and assuming $c_n / c_\text{s,max}$ is equivalent to the local SoC $z_n$ in \cref{eq:diff_zn}, we obtain
\begin{align} \label{eq:discretized2}
	\frac{\dd{z_n}}{\dd{t}} \Omega_n =
	\frac{z_{n+1} - z_{n}}{b/\qty(S_{n}D_\text{s})} - \frac{z_n - z_{n-1}}{b/\qty(S_{n-1}D_\text{s})}.
\end{align}
Substituting \cref{eq:discretized2} into \cref{eq:diff_zn2} and then comparing with \cref{eq:shelleq2}, we relate the fitted diffusion resistance to the diffusivity: 
\begin{align} \label{eq:resistance2}
	R_{\text{d},n} = \frac{kb}{F c_\text{s,max} N_\text{a}S_{n} D_\text{s}}. 
\end{align}
Following the same procedures as in \cref{eq:resistance,eq:outerres,eq:timescale} leads to \cref{eq:timescale-fit} and the expression of diffusivity in terms of the fitted resistance:
\begin{align}
	D_\text{s} = \frac{kN}{4\pi F c_\text{s,max} N_\text{a} a R_{\text{d},1}}. 
\end{align}

\addcontentsline{toc}{section}{References}
\bibliography{./manuscript.bib}


\end{document}